# Digitizing scientific data and data retrieval techniques

Ranjeet Devarakonda, Giri Palanisamy, Jim Green

May 2010
Oak Ridge National Laboratory
Oak Ridge, TN 37831

Storing data is easy, but finding and using data is not. It is desirable that the data is stored in a structured format, which can be preserved and retrieved in future. Creating Metadata for the data is one way of creating structured data formats. Metadata can provide Multidisciplinary data access and will foster more robust scientific discoveries. In the recent years, there has been significant advancement in the areas of scientific data management and retrieval techniques, particularly in terms of standards and protocols for archiving data and metadata. New search technologies are being implemented around these protocols, which makes searching easy, fast and yet robust. Scientific data is generally rich, not easy to understand, and spread across different places. In order to integrate these pieces together, a data archive and an associated metadata is generated. This data should be stored in a format that can be locatable, retrievable and understandable, more importantly it should be in a form that will continue to be accessible as technology changes, such as XML.

Scientific data, in its most general context across multiple disciplines, includes measurements and observations of natural phenomena for the purpose of explaining the behavior of or testing hypotheses about the natural systems. This includes observational data captured in real-time by sensors, surveys, and imaging devices; experimental data from laboratory instruments, for example, gene sequences, chromatograms, and characterization of samples; simulation data generated from models where the model is equally important with the input and output data such as for climate and economic models; and derived or compiled data that are the result of text and data mining, and compiled and integrated databases from multiple sources.

Remote sensing and high throughput scientific instruments and high spatial and temporal resolution model simulations are creating vast data repositories and as a result of increased investments in engineering research, disease research , and environmental monitoring, the volume of scientific data is approximately doubling each year.  (Gray et al., 2005. Technical Report MSR-TR-2005-10).  Data volume requires new scientific methods to analyze and organize the data for fully exploiting the value of the data and beyond the original collection purpose.

The data content can be collected in any format, most is digital by necessity and to be of greatest value for analyses. Instrument platform and science program specific formats are often integrated into the data processing stream. The data can be formatted as straightforward tabular ASCII files, spatially gridded files, Geotiff image formats or formatted to be consistent with programmatic repository needs such as Genomes-to-Life.

The digital data are not complete without descriptive information (meta)data to tell us what it means. Metadata are the descriptive information about data that explains the measured attributes, their names, units, precision, accuracy, data layout and ideally a great deal more. Most importantly, metadata includes the data lineage that describes how the data was measured, acquired, or computed (Gray et al., 2005. Technical Report MSR-TR-2005-10). Metadata enables data sharing and access. Sharing data with colleagues the broader scientific community and public is highly desirable and will result in greater advancement of science.

Mercury, a metadata harvesting, data discovery, and access system, built for researchers to search for, share and obtain spatiotemporal data used across a range of climate and ecological sciences. Mercury harvests the structured metadata and key data from several data providing servers around the world and builds a centralized index. Mercury enables scientists to search for terrestrial ecology data located in diverse NASA and non-NASA data centers, using a straightforward web browser interface, and integrating multiple different types of metadata.